\documentclass[aps,prb,reprint,superscriptaddress]{revtex4-1}
\usepackage{bbold}
\usepackage{graphicx}% Include figure files
\usepackage{dcolumn}% Align table columns on decimal point
\usepackage{bm}% bold math
\usepackage{epstopdf}
\usepackage{ulem}
\usepackage{color}
\usepackage{longtable} 
\usepackage{ulem}
\newcommand{\yb}{Yb$^{3+}$}

\newcommand{\dfc}{$^2$F$_{5/2}$}
\newcommand{\dfs}{$^2$F$_{7/2}$}

\newcommand{\pr}{Pr$^{3+}$}

\newcommand{\tm}{Tm$^{3+}$}

\newcommand{\nd}{Nd$^{3+}$}
\newcommand{\er}{Er$^{3+}$}

\newcommand{\eu}{Eu$^{3+}$}

\newcommand{\ce}{Ce$^{3+}$}

\newcommand{\YSO}{Y$_2$SiO$_5$}
\newcommand{\YAG}{Y$_3$Al$_5$O$_{12}$}

\newcommand{\LNO}{LiNbO$_3$}
\newcommand{\YVO}{YVO$_4$}

\newcommand{\wnm}{cm$^{-1}$}

\newcommand{\yt}{Y$^{3+}$}

\newcommand{\SW}{\textcolor{black}}

\begin{document}

% Use the \preprint command to place your local institutional report
% number in the upper righthand corner of the title page in preprint mode.
% Multiple \preprint commands are allowed.
% Use the 'preprintnumbers' class option to override journal defaults
% to display numbers if necessary
%\preprint{}

%Title of paper
\title{High Resolution Optical Spectroscopy and Magnetic Properties of Yb$^{3+}$ in  Y$_2$SiO$_5$}

% repeat the \author .. \affiliation  etc. as needed
% \email, \thanks, \homepage, \altaffiliation all apply to the current
% author. Explanatory text should go in the []'s, actual e-mail
% address or url should go in the {}'s for \email and \homepage.
% Please use the appropriate macro foreach each type of information

% \affiliation command applies to all authors since the last
% \affiliation command. The \affiliation command should follow the
% other information
% \affiliation can be followed by \email, \homepage, \thanks as well.
\author{Sacha Welinski}
%\email[]{Your e-mail address}
%\homepage[]{Your web page}
%\thanks{}
%\altaffiliation{}
\affiliation{PSL Research University, Chimie ParisTech, CNRS, Institut de Recherche de Chimie Paris, 75005, Paris, France}

\author{Alban Ferrier}
\affiliation{PSL Research University, Chimie ParisTech, CNRS, Institut de Recherche de Chimie Paris, 75005, Paris, France}
\affiliation{Sorbonne Universit\'es, UPMC Univ Paris 06, Paris 75005, France}

\author{Mikael Afzelius}
\affiliation{Group of Applied Physics, University of Geneva, CH-1211 Geneva 4, Switzerland}

\author{Philippe Goldner}
\email{philippe.goldner@chimie-paristech.fr}
\affiliation{PSL Research University, Chimie ParisTech, CNRS, Institut de Recherche de Chimie Paris, 75005, Paris, France}

%Collaboration name if desired (requires use of superscriptaddress
%option in \documentclass). \noaffiliation is required (may also be
%used with the \author command).
%\collaboration can be followed by \email, \homepage, \thanks as well.
%\collaboration{}
%\noaffiliation

\date{\today}

\begin{abstract}
Rare earth doped crystals are promising systems for quantum information processing. In particular paramagnetic rare earth\SW{s} could be used to build coherent interfaces with optical and microwave photons. In addition, isotopes with non zero nuclear spins could provide long lived states for quantum state storage and processing. \yb{} is particularly interesting in this respect since it is the only paramagnetic rare earth with a spin 1/2 isotope, which corresponds to  the simplest possible level structure. In this paper, we report on the optical and magnetic properties of \yb{} in the two sites of  \YSO \SW{, a commonly used crystal for quantum applications}.  We measured  optical inhomogeneous linewidths, peak absorption coefficients, oscillator strengths, excited state lifetimes and \SW{fluorescence} branching ratios. 
%The values found are among the best for rare earth doped materials. 
The Zeeman tensors were also determined in the ground and excited states, as well as the ground state hyperfine tensor for the $^{171}$\yb{} ($I=1/2$) isotope. These results suggest that \yb:\YSO{} is a promising material for applications like solid state optical and microwave quantum memories.

\end{abstract}

% insert suggested PACS numbers in braces on next line
\pacs{}
% insert suggested keywords - APS authors don't need to do this
%\keywords{}

%\maketitle must follow title, authors, abstract, \pacs, and \keywords
\maketitle

% body of paper here - Use proper section commands
% References should be done using the \cite, \ref, and \label commands
\section{Introduction}
\label{intro}

Rare-earth (RE) doped crystals are promising solid state candidates for quantum information processing. \cite{Tittel:2010bp,Thiel:2011eu,Goldner:2015ve,deRiedmatten:2015tj} In particular, they can show extremely narrow optical homogeneous linewidths at low temperature, in the range  of a few kHz to less than 100 Hz.\cite{Macfarlane:2002ug,Bottger:2009ik} Combined with inhomogeneous linewidths in the GHz range, this allows optical addressing of ions within ensemble for quantum processors, or quantum memories with large time-bandwidth products. Moreover, several RE ions have non-zero nuclear spins, which can act as long-lived, optically addressable qubits. \cite{Louchet:2008cj,Fraval:2005gk} As an example, hyperfine transitions of \eu:\YSO{} can show coherence lifetimes ($T_2$)  up to 6 hours at 2 K [\onlinecite{Zhong:2015bw}] using various dephasing control techniques,  and still reach a few ms at 20 K.\cite{Arcangeli:2015hd} Recent results include entanglement storage\cite{Clausen:2011uw,Saglamyurek:2015esa}  and light-matter teleportation at the telecom wavelength,\cite{Bussieres:2014dc} single photon \SW{level} memories  with storage in nuclear spin states,\cite{Jobez:2015gt,Gundogan:2015dh} as well as memories with high efficiency,\cite{Hedges:2010dq,Dajczgewand:2014et}  and storage time exceeding one minute.\cite{Heinze:2013co} Large and switchable interactions between RE ions have also been observed,\cite{Ahlefeldt:2013jl} as well as  single RE detection,\cite{Xia:2012gu,Utikal:2014bd} which opens the way to quantum processing in these systems.

Paramagnetic RE ions, \SW{such as} \nd{} or \er{}, have one more degree of freedom \SW{due to their} electron spins. This can be used to interface microwave photons to a RE doped crystal through a superconducting resonator\cite{Staudt:2012he,Probst:2013hn} and obtain quantum memories for superconducting qubits. \cite{Afzelius:2013ga,Probst:2015ku,Arcangeli:2016eu} In this case too, nuclear \SW{hyperfine} transitions can  provide long storage time. We recently showed that $^{145}$\nd:\YSO{} hyperfine transitions have ground state coherence lifetimes up to 9 ms at 5 K, whereas the electron spin $T_2$ is about 100 $\mu$s.\cite{Wolfowicz:2015ex} High fidelity coherent transfer of microwave excitations to nuclear spins was also demonstrated by quantum state tomography. Ultimately, it should be possible to build  coherent interfaces between optical and/or microwave photonics qubits, and long lived nuclear spin quantum states. 

\yb{} ions have attractive properties in this respect. Their  4f$^{13}$ configuration comprises only two multiplets: \dfc{} (ground) and \dfs{} (excited) separated by $\approx 10000$ \wnm. This energy corresponds to the near infrared, \SW{where laser diodes are easily found.}
The excited state decays usually radiatively, with a lifetime of about 1 ms, which sets a limit for the optical coherence lifetime of 2 ms. There are also two naturally abundant isotopes  with non-zero nuclear spin: $^{171}$\yb ($I=1/2$) and $^{173}$\yb ($I=5/2$), which hyperfine transitions could be used as long lived qubits. The lower spin angular momenta of \yb{} compared to \er{} and \nd{} ($I=7/2$ for all non-zero spin isotopes) \SW{result} in simpler energy level structures. This is important since optical initialization and coherent manipulation of spins require selecting specific transitions within the optical inhomogeneous linewidth by potentially complex optical pumping sequences. The lower number of spin states in $^{171}$\yb{} and $^{173}$\yb{} could simplify them considerably.

Here we report on the low temperature and high resolution optical spectroscopy of \yb:\YSO. This crystalline host has been chosen since it shows outstanding properties in terms of narrow optical linewidths and long coherence spin lifetimes when doped with \pr, \nd,\eu and \er.\cite{Macfarlane:2002ug,Usmani:2010hd,Wolfowicz:2015ex} It is currently the most used host in quantum storage  experiments. 
We measured inhomogeneous linewidths, absorption spectra, excited state lifetimes and branching ratios for the two \yb{} sites and found  values among the best for RE doped crystals.  Magnetic fields are often used to slow down electron spin relaxation,\cite{Bottger:2009ik} which influences optical and spin coherence lifetimes and spectral hole burning efficiency.  As a first step towards these dynamical experiments, we  determined the ground and excited state $g$ tensors from electron paramagnetic resonance and optical spectroscopy.  Ground state hyperfine tensors were also found for  $^{171}$\yb{} and $^{173}$\yb, giving   hyperfine structures extrapolated at low magnetic field of a few GHz, suitable for coupling to superconducting resonators. 
 Overall, these results suggest that \yb:\YSO{} is suitable to build coherent interfaces with optical and microwave photons, while offering  nuclear spin transitions for long lived quantum states.

%\begin{itemize}
%\item RE paramagnetic ions:
%	\begin{itemize}
%	\item coupling to superconducting qubits
%	\item fast control with electron spin
%	\item nuclear spins provide the long term memory (and also can provide larger splittings at low fields)
%	\item provide also optical access
%	\end{itemize}
%\item Yb3+
%	\begin{itemize}
%	\item convenient transition in the IR accessible by diodes
%	\item low nuclear spin compared to Nd and Er, may facilitate SHB
%	\item never studied before, we need basic information which are in fact key parameters
%	\item in particular low symmetry implies that tensors must be determined or absorption in different directions, complex system
%	\item YSO because low spin density and very successful with other RE ions
%	\end{itemize}
%\item here
%	\begin {itemize}
%	\item optical properties inhomogeneous linewidths, radiative lifetime, peak absorption coefficients, oscillator strengths polarized etc? 
%	\item g tensors for ground and excited states and hyperfine ground tensors useful for coupling but also for reducing decoherence and spectral diffusion effects
%	\item preliminary results show that holes of 1 s lifetimes can be burned
%	\item Yb is very promising for memories
%	\end{itemize}
%\end{itemize}

\section{Experimental}

We used \YSO{} (YSO) samples %yttrium orthosilicate crystals ($\text{Y}_{\text{2}}\text{SiO}_{\text{5}}$)
 doped at 0.005 at.\% (50 ppm) with $\text{Yb}^{\text{3+}}$  and cut from a boule   grown by the Czochralski method. YSO has a monoclinic structure belonging to the $C^{6}_{2h}$ ($C_{2}/c$) space group. $\text{Yb}^{\text{3+}}$  can substitute \yt{}  in  two different crystallographic sites, both with a $C_{1}$ point symmetry. \SW{In addition, non equivalent subsites appear under magnetic fields neither parallel nor perpendicular to $b$. They are related by the crystal $C_2$ ($b$) symmetry axis.} 
% The three crystallographic axis of YSO are communally labeled $a$, $b$ and $c$. The b axis has a $C_{2}$ symmetry axis. $a$ and $c$ are perpendicular to $b$ and the angle between them is $\beta=102^\circ 39'$. 
 Samples  were cut along the three principal dielectric axes: $b$ (the $C_2$ crystallographic axis), $D_{1}$ and $D_{2}$. 
% For optical studies, two samples with different sizes (6.36, 4.77, 5.06) and (7.14, 1.41, 5.97) mm in the ($b$, $D_{1}$, $D_{2}$) frame were used to avoid too large absorption. For the Electron Paramagnetic Resonance study the size of the sample is (1.56, 2.23, 2.02) millimeters.\\
%\indent{}The $\text{Yb}^{\text{3+}}$ ions can substitute $\text{Y}^{\text{3+}}$ in both sites equally in $\text{Y}_{\text{2}}\text{SiO}_{\text{5}}$. Since those two sites have different coordination with oxygen atoms, the crystal field levels of $\text{Yb}^{\text{3+}}$ in the two sites are not the same. 
%Moreover, each site possesses two sub-sites chemically equivalent but not magnetically equivalent except in the $D_{1}D_{2}$ plane.
Ytterbium has five stable even isotopes $^{\text{168}}\text{Yb}$, $^{\text{170}}\text{Yb}$, $^{\text{172}}\text{Yb}$, $^{\text{174}}\text{Yb}$ and $^{\text{176}}\text{Yb}$ with nuclear spin I=0 and a total  abundance of 69.59\%. There are also two odd isotopes, $^{\text{171}}\text{Yb}$, with I=1/2 and an abundance of 14.28\% and $^{\text{173}}\text{Yb}$ with I=5/2 and a abundance of 16.13\%.

Absorption spectra with 0.1 nm resolution were obtained with a Varian Cary 6000i \SW{spectrophotometer}. Fluorescence measurements were performed using a Coherent 829 Titanium Sapphire laser pumped by a Coherent Verdi G10 laser, a SpectraPro 750 monochromator (1 nm resolution) and an InGaAs photodiode. \SW{Fluorescence} lifetimes were measured with a tunable optical parametric oscillator pumped by a Nd:YAG laser (Ekspla NT342B-SH,  6 ns pulse length) as the excitation source, a Jobin-Yvon HR250 monochromator and an InGaAs photodiode. 
High resolution transmission spectra were recorded by scanning  a single mode Toptica DL 100 diode laser (1 MHz linewidth)  around 980 nm. Continuous frequency scans of about 15 GHz could be  performed.   The laser beam  was collimated with a power of about 1mW in front of the cryostat. The transmitted signal was detected by a Thorlabs PDA36A photodiode and a reference beam by a Thorlabs PM10A photodiode. A small part of signal was also sent to a Toptica Fabry-Perot Interferometer (1 GHz Free Spectral Range at 980 nm).  This allowed us to precisely calibrate  the spectra frequency scale. The sample was maintained at 10 K in a CTI-Cryogenics closed-cycle cryostat.

Electron paramagnetic resonance spectra were recorded at 9 K with a Bruker ELEXSYS E500 and an ELEXSYS Super High Sensitivity Probe Head in X-band. 
For the optical determination of $\text{Yb}^{\text{3+}}$ excited state g-tensors, the crystal was placed between two permanent NdFeB magnets. The field average value was 217 mT with an  inhomogeneity of about 10 \% along the laser propagation axis. To record angular variations, the crystal sat on a pedestal attached to an Attocube ANRv51 stage and was rotated by steps of $10^\circ$. The whole assembly was put in a Janis LHe  cryostat at 10 K. The transmission spectra were recorded with the set up described above. 

%\indent{}Spectral holeburning measurements were also performed using the Attocube device with a permanent magnetic field of about 1600G. The measurements were made at several temperature between 4K and 1.5K. An acousto-optic modulator monitored by an arbitrary waveform generator Agilent N8242A was set up to create the burning and scanning pulse sequences.

\section{Results and Discussion}

\subsection{Optical Spectroscopy}
\label{optSpec}

\subsubsection{Absorption and Emission Spectra}
\label{absEm}
\begin{figure}
\includegraphics[width=0.85\columnwidth]{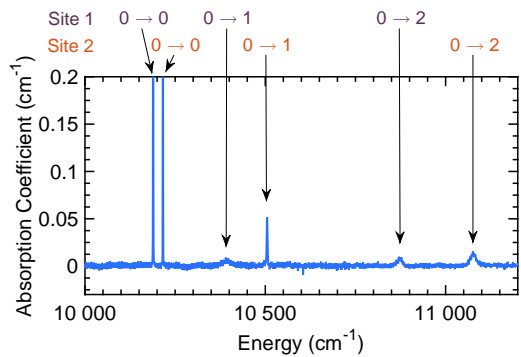}
\caption{Unpolarized absorption spectrum of \yb:\YSO{} at 12 K with light propagating along the $b$ axis. Transitions between CF levels are indicated for sites 1 and 2.}
\label{abs12K}
\end{figure}

We first recorded absorption  and emission spectra to determine \yb{} crystal field (CF) level energies. The absorption spectrum (Fig. \ref{abs12K}) shows well \SW{resolved} lines corresponding to transitions from the lowest CF level \dfs{}(0) of the ground multiplet  to the three CF levels \dfc{}(0,1,2) of the excited one. These lines are homogeneously broadened except for the two lowest energy ones \SW{which are likely to be inhomogeneously broadened. \cite{Kis:2014fq}}  A narrow and isolated peak, with a full width at half maximum (FWHM) of 2.8 \wnm, is recorded at 10505 \wnm{} and corresponds to the \dfs(0)$\rightarrow$\dfc(1) transition for site 2. Other transitions above 10300 \wnm{} are much broader with  FWHM between 15 and 25 \wnm. The structures in the range 10450 -10650 \wnm{} observed for  \yb{} concentrations of a few \% [\onlinecite{Gaume:2002vk}] are not seen in this 50 ppm doped sample and are therefore attributed to distorted \yb{} sites. 

\begin{figure}
\includegraphics[width=0.85\columnwidth]{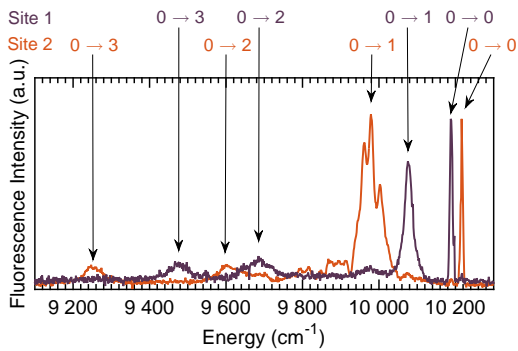}
\caption{Emission spectra of sites 1 (purple line) and 2 (orange line) of \yb:\YSO{} at 10 K excited respectively at 919 nm and 902 nm. Transitions between CF levels are indicated.} 
\label{em10K}
\end{figure}

Low doping concentration also prevents energy transfer between sites and allows separate recording of each site's emission spectrum (Fig. \ref{em10K}). These lines correspond to the \dfc(0)$\rightarrow$\dfs(0,1,2,3) transitions and, except for the higher energy lines, \SW{whose} widths are instrument limited, are much broader than those observed in absorption. In addition, many partially resolved lines, that we attribute to vibronic transitions,  appear in the range 9760-10060 \wnm{} for site 2.  This induces an uncertainty in the \dfs(2) level position, which we determined using the  strongest peak  at 9982 \wnm.

The crystal field levels for both sites are summarized in Table \ref{optTable} and are in good agreement with previous studies.  \cite{Gaume:2002vk,Denoyer:2008kx} CF splittings for the ground and excited multiplets are significantly smaller for site 1, which suggests that it corresponds \SW{to} the crystallographic site with a coordination number (CN) of 7, in which Y-O distances are larger. Indeed, scalar CF strengths  have been found smaller for this site in \ce:\YSO{} in a theoretical study. \cite{Wen:2014eh}
In the same way, \pr{} ions with the smaller CF splittings have also a higher absorption coefficient, suggesting a higher relative concentration.\cite{Equall:1995wt} Since  \pr{} ions have a larger ionic radius than \yt{} ones, they should occupy preferentially the CN = 7 site, which has a larger volume.

\begin{figure}
\includegraphics[width=0.85\columnwidth]{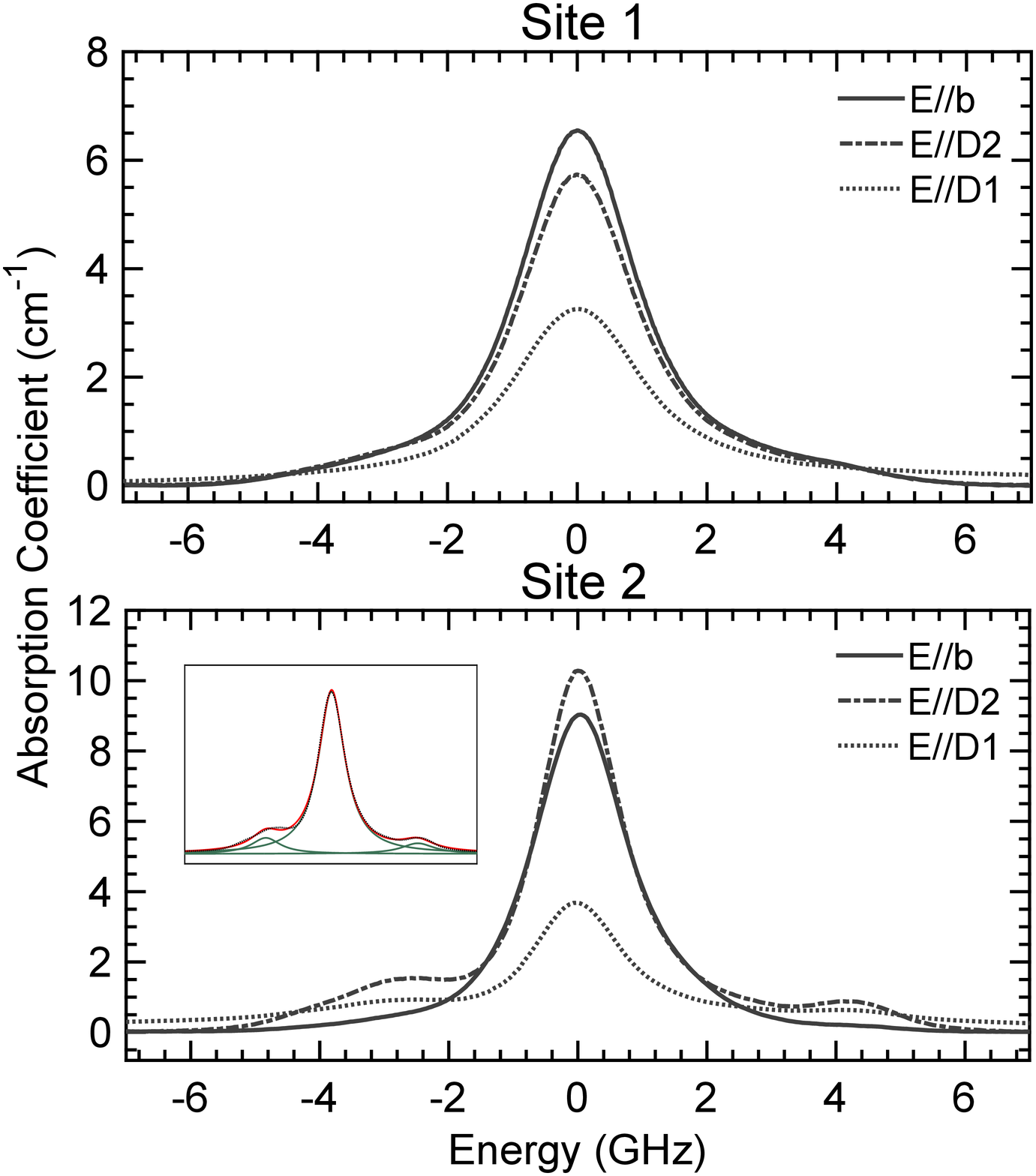}
\caption{High resolution absorption spectra of sites 1 (upper graph) and 2 (lower graph) in \yb:\YSO{} at 10 K for light electric field polarized along $b$, $D_1$ and $D_2$ axes. Inset: deconvolution of site 2 spectrum for  $E\parallel D_2$ by three Lorentzian curves.}
\label{hiRes}
\end{figure}

\begin{table*}%[H] add [H] placement to break table across pages
 \caption{\label{optTable} Optical transitions in \yb:\YSO. CF level energies; 0-0 transition wavelength (vac. ), inhomogeneous linewidths, peak absorption coefficient for different light polarizations, oscillator strength; experimental fluorescence decay time ($T_1$), spontaneous emission decay time for the 0-0 transition ($T_{1s}$) and $T_{1s}/T_1$ ratio.}
\begin{ruledtabular}
 \begin{tabular}{ccccccccccc}
         & \multicolumn{2}{c}{Energy (\wnm)} & $\lambda_{vac}$ (nm) & $\Gamma_\mathrm{inh}$ (GHz) & $\alpha_0$ (\wnm ) & $P \times 10^7$ & $T_1$ (ms) & $T_{1s}$ (ms) & $T_{1s}/T_1$\\
& \dfs & \dfc \\        
Site 1 & 0 & 10189 & 981.463&2.2& 3.3 ($E \parallel D_1$)  & 5.0 & 0.87 & 8.5 & 9.8 \\
        & 111 & 10391 && &5.7 ($E \parallel D_2$) & \\
        & 499 &  10874 && &6.5 ($E \parallel b$) & \\
        & 709 \\
Site 2  &   0 & 10216 & 978.854&1.7 & 3.7 ($E \parallel D_1$) & 6.4& 1.3 & 6.5 & 5.0\\
	  & 234 & 10505 & &&10.3 ($E \parallel D_2$) &\\
	  & 612 & 11076 & &&9.0 ($E \parallel b$) &\\
	  & 970
 \end{tabular}
\end{ruledtabular}
 \end{table*}
Because of phonon relaxation between CF levels within the multiplets, only the 0-0 (\dfs(0)-\dfc(0)) transition can have a narrow homogeneous linewidth at low temperatures.   
It is therefore the transition of interest for QIP applications, \cite{Goldner:2015ve} and we focus on it in the following. 
High resolution absorption spectra for the 0-0 transitions were obtained at zero external magnetic field by scanning a single mode diode laser. \SW{Spectra were obtained for the light electric field $E$ polarized along the $D_2$  and $b$ axes and the wave vector $k$ parallel to $D_1$; for $E$ parallel to  $D_1$, $k$ was parallel to $D_2$ (Fig. \ref{hiRes})}. The lines peak at 10188.87 \wnm (981.463 nm in vac.) and 10216.06 \wnm (978.854 nm in vac.) for sites 1 and 2. 

Both lines show a narrow central part, as expected for the $I=0$ \yb{} isotopes, on top of a broader and weaker structure, in which peaks can be  clearly seen in some cases (e.g. site 2, $E \parallel D_2$).  We attribute this additional feature to the zero field hyperfine structures of the  $^{171}$\yb{}(abundance 14.3 \%)  and $^{173}$\yb{} (abundance 16.1 \%) isotopes, which span about 3-4 GHz in the ground state (see Section \ref{magProp}). All lines could be well fitted by a combination of Lorentzian lines (see inset in Fig. \ref{hiRes}). This has been observed in several RE doped crystals \cite{Konz:2003kw,Beaudoux:2012ch} and, according to Stoneham \cite{Stoneham:1969wq}, corresponds to perturbations by a low concentration of point defects. The FWHM of the central parts of the lines are 2.2 and 1.7 GHz for sites 1 and 2. These values are comparable to those found for \YSO{} doped at low levels of  RE ions \cite{Macfarlane:2002ug} and are related to the difference in ionic radius between \yt{}(0.892 \AA) and \yb{} (0.858 \AA). \cite{Shannon:1969dj} It is important to note that growth conditions can also have a significant influence on $\Gamma_\mathrm{inh}$ [\onlinecite{Ferrier:2015ku,Bottger:2006dg}].

Large and anisotropic peak absorption coefficients were measured, reaching maximum values of 6.5  \wnm ($E \parallel b$) and 10.3 \wnm ($E \parallel D_2$) for site 1 and  2 respectively. In both cases, the lower absorption occurs for a light electric field polarized along $D_1$.    The average oscillator strengths $P$ were calculated without local field corrections
%\begin{equation}
%P = \frac{m c^2}{\pi N e^2} \int \alpha(\nu) d\nu}
%\end{equation}
%
%
 and assuming equal \yb{} occupancy for the two sites, because of the close ionic radii of \yt{} and \yb. Site 2 value ($P=6.4 \times 10^{-7}$) is about 30\% higher than the one for site 1 ($P=5.0 \times 10^{-7}$). 
 
 Excited state population lifetimes are around 1 ms for both sites (Table \ref{optTable}) and can be considered to be purely radiative. On one hand the low doping concentration prevents energy transfers between \yb{} ions  and  to quenching centers. On the other hand, the energy gap between the \dfs{} and \dfc{} multiplets ($\approx$ 10000 \wnm) is much larger than the \YSO{} phonon cut-off frequency ($\approx  960$ \wnm [\onlinecite{Campos:2004cv}])  so that multiphonon relaxation is negligible.

\subsubsection{Comparison with other Rare Earth Ions}

In the context of quantum processors and quantum memories, RE with strong optical transitions are needed since they can be coherently driven at a faster rate and result in larger opacity. \cite{Goldner:2015ve} Another important parameter for RE coupled to cavities is the strength of the transition of interest and its relation to all other possible transitions. \cite{McAuslan:2009er}

 As the cavity enhances only resonant transitions, achievable emission rates with Purcell enhancement are proportional to the emission rate through the transition that would be coupled to the cavity, 0-0 in our case.
The corresponding radiative lifetime $T_{1s}$ is expressed as:
\begin{equation}
\left (T_{1s}\right)^{-1} = \frac{2\pi e^2n^2 \nu^2}{\epsilon_0 m c^3}P
\end{equation}
where $e$, $m$ are the electron charge and mass, $\nu$ is the transition frequency, $n$ the refractive index, $c$ the speed of light, and $\epsilon_0$  the vacuum permittivity. $T_{1s}$ should be as short as possible, with $T_1$ as a lower bound. On the other hand, decays trough  uncoupled transitions or non-radiative paths represent a loss for the cavity-emitter system and should be minimized. The ratio $T_{1s}/T_1$ has therefore to be as small as possible.
As showed in Table \ref{optTable}, site 2 presents the best values  for $P$, $T_{1s}$ and $T_{1s}/T_1$. 

Compared to other RE doped crystals investigated for QIP, its oscillator strength is \SW{one of} the highest reported in the YSO host.\cite{McAuslan:2009er,Goldner:2015ve} It is more than twice the value found in \pr{}, three times that of \er{} and 5 times that of \eu. This is mainly due to the 0-0 transition originating from the first excited multiplet and the low $J$ value of the ground multiplet \SW{of \yb{}}.  Combined with a  relatively short and purely radiative $T_1$, the reduced number of  possible transitions favors a strong 0-0 transition. Still, some RE-host combinations allow larger oscillator strengths, like \nd:\YVO{} ($8 \times 10^{-6}$) or \pr:\YAG{} ($1.5 \times 10^{-6}$).\cite{McAuslan:2009er}

$T_{1s}$ is also short, similar to what is found in \pr:YSO (5.7 ms) and much shorter than for \er{} (54.6 ms) or \eu{} (120 ms) doped YSO. %\SW{\sout{In this respect, one additional advantage of \yb:YSO emission is its near infrared wavelength, for which high efficient single photon detectors are available.}}
Compared to systems with very high oscillator strengths, \SW{such as} \nd:\YVO,  $T_{1s}$ is however about 13 times longer. Finally,  $T_{1s}/T_1$ is very close to the lowest value observed for YSO, 4.8 in \er:YSO and about one order of magnitude better than for \pr{} (34.5) and \eu{} (63.2). Only \tm:\LNO{} shows a significantly lower ratio (2.25). 
In summary, the 0-0 transition of \yb{} ions in site 2 of YSO  shows  among the best properties for RE doped crystals in terms of oscillator strength, spontaneous decay $T_{1s}$ and $T_{1s}/T_1$ ratio.

\subsection{Magnetic Properties}
\label{magProp}

\begin{figure}
\includegraphics[width=0.85\columnwidth]{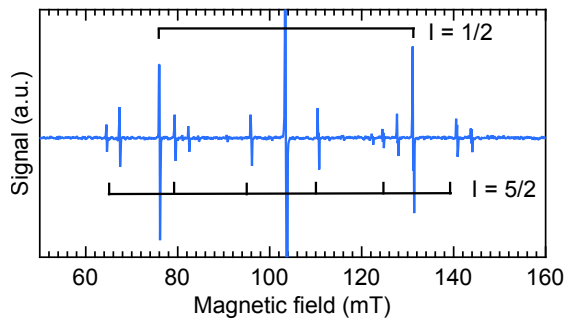}
\caption{Site 1 EPR spectrum obtained for $B$ at $160^\circ$ from $D_1$ in $D_1D_2$ plane in \yb:\YSO. Lines corresponding to the $I=0, 1/2$ and $5/2$ isotopes are observed (see text) and calculated from a spin Hamiltonian model with no quadrupolar contribution.} 
\label{EPRspec}
\end{figure}

We first determined \yb{} ground state Zeeman and hyperfine tensors by EPR, \SW{complementing previous studies where only particular orientations of the magnetic field were investigated.\cite{Kurkin:1980th,Denoyer:2008kx}} Due to the two crystallographic sites, which generally divide in two sub-sites under a magnetic field, and the two \yb{} isotopes with $I\neq 0$, many lines were observed for magnetic fields $B$ in the range 50-1000 mT. Fig. \ref{EPRspec} shows site 1 lines for $B$ at $160^\circ$ from $D_1$ in $D_1D_2$ plane, a configuration where the two subsites, related by a $C_2$ symmetry along $b$, are equivalent. Apart from the Zeeman line at 103.5 mT, the two transitions corresponding to $^{171}$\yb{} ($I=1/2$) are observed at 76.1 and 131.2 mT. The other lines located between 64.5 and 144.6 mT are attributed to the hyperfine structure of $^{173}$\yb{}($I=5/2$). In this case, due to quadrupole interactions, some transitions with $\Delta M_I\neq 0$ are clearly seen. For comparison, the positions of lines corresponding to $\Delta M_I=0$ and deduced from $^{171}$\yb{} hyperfine tensor (see below) are indicated. 
The lines were very narrow at low magnetic field and \SW{FWHMs} as low as 12 MHz could be recorded, which is favorable to coupling with high quality factor microwave resonators. \cite{Probst:2013hn} 

\begin{figure}
\includegraphics[width=0.85\columnwidth]{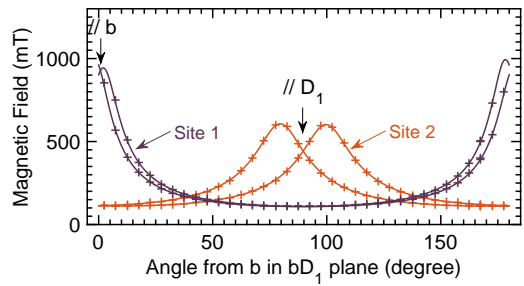}
\caption{ Angular variation of EPR transitions in the $bD_1$ plane ($I=0$ isotopes) for sites 1 and 2 in \yb:\YSO. Crosses: experimental data; lines:  fit to a spin Hamiltonian model.} 
\label{EPRvar}
\end{figure}

The $\mathbf{g}$ tensors of the $I=0$ isotopes were determined for both sites from angular variations of the Zeeman lines in the three perpendicular planes $bD_1$, $bD_2$ and $D_1D_2$ (Figs. \ref{EPRvar} and S1). All line positions were simultaneously fitted to the Hamiltonian $H=\mu_B \mathbf{B}\cdot\mathbf{g}\cdot\mathbf{S}$, where $\mu_B$ is the Bohr magneton and $\mathbf{S}$ is an effective 1/2 spin operator. The $C_2$ symmetry linking the subsites was also taken into account. The effective spin approach is possible because of the large ground state CF splittings (100 - 200 \wnm) compared to the Zeeman one (0.3 \wnm).  Moreover, at low temperature, only the lowest energy CF doublet is populated. For each plane, misalignment  between the crystal and the lab frame was also allowed for by introducing  three variable angles in the fit. Excellent agreement with experimental data was obtained, as shown in Figs. \ref{EPRvar} and S1. The principal  values of $\mathbf{g}$ and  the Euler angles for the principal axes are gathered in Table \ref{magTable}.

\begin{figure}
\includegraphics[width=0.9\columnwidth]{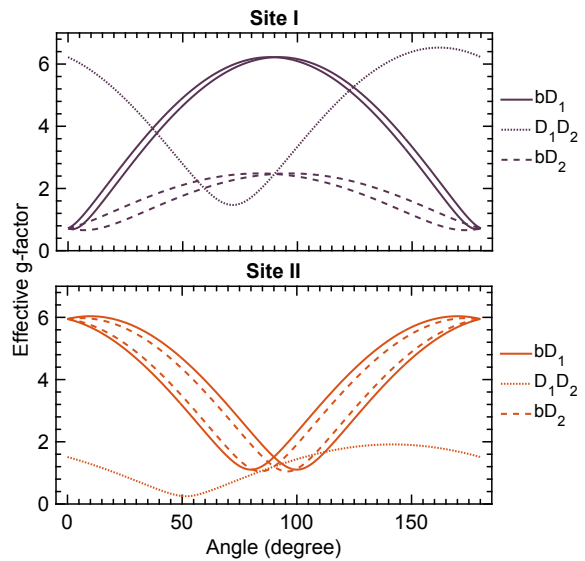}
\caption{Calculated effective $g$ factors in the $bD_1$, $bD_2$ and $D_1D_2$ planes for sites 1 and 2 in \yb:\YSO. } 
\label{gFact}
\end{figure}

Fig. \ref{gFact} shows the \SW{calculated} effective $g$ factors for magnetic fields oriented in the $bD_1$, $bD_2$ and $D_1D_2$ planes. For site 1,  the largest principal value is 6.53, with the principal axis oriented close to the $D_1$ axis.  For site 2, the largest principal value is similar, 6.06, but corresponds to an orientation close to the $b$ axis. This explains the main features of the angular variations of Fig. \ref{EPRvar}. For both sites, the two other principal values are much smaller. This results in low $g$ factors in the $bD_2$ and $D_1D_2$ planes for sites 1 and 2 respectively. 

As mentioned above, EPR lines were fitted allowing for crystal misalignment in each plane. For site 1, misalignment angles were below 2$^\circ$. Surprisingly, these angles were much larger in the case of site 2, reaching for example 7$^\circ$ for the $D_1D_2$ variation. Moreover,  fitting site 1 transitions with site 2 angles  resulted in poor agreement with  experimental data. This rules out a higher sensitivity of site \SW{2} transitions to crystal misalignment. This discrepancy is  particularly clear in the $D_1D_2$ angular variation (Fig. S1). For magnetic fields around 412 mT (but for different crystal orientations), subsite lines are separated by only 1.2 mT for site 1,  whereas site 2 lines are separated by  21.3 mT. This result suggests that the  distortion introduced by  \yb{} substituting \yt{} in site 2 results in an effective $C_2$ axis direction significantly different from the host crystal. \SW{This observation should be further investigated, as it is of importance for predicting particular energy level structures, such as Zero First Order Zeeman shift (ZEFOZ) lines. \cite{Fraval:2004cu,McAuslan:2012eb}}
% This would be in agreement with site 2 corresponding to the  \yt{} site with the largest volume. \yb being smaller than \yt, this site would be more distorted upon substitution. 

%The fitted values did not exceed 10 degrees, consistent with the accuracy of the cutting and positionning of small samples. 

\begin{figure}
\includegraphics[width=0.9\columnwidth]{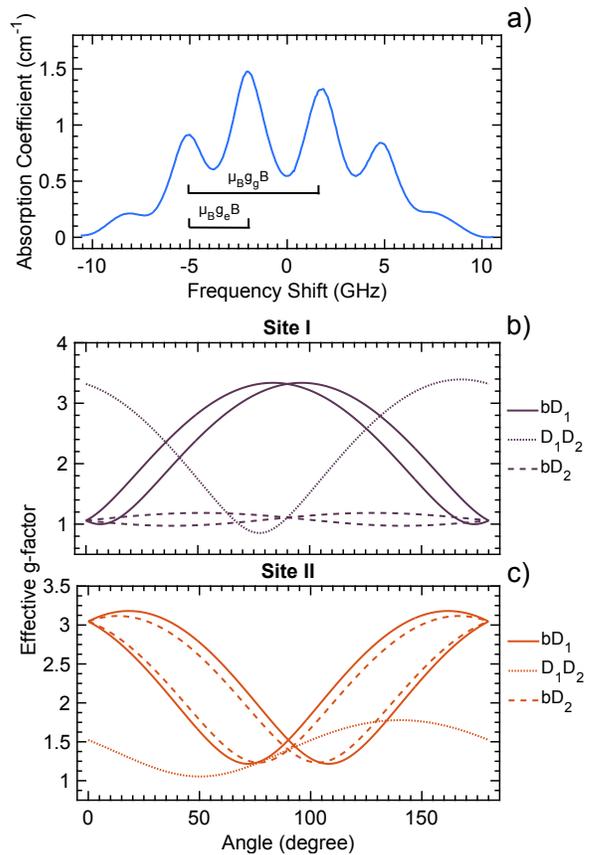}
\caption{(a) Transmission spectrum of site 1 in \yb:\YSO{} under a magnetic field of 217 mT along the $D_2$ axis ($T  = 10$ K). Energy separations corresponding to the ground and excited state effective $g$ factors are indicated. (b) and (c) calculated excited state effective $g$ factors in the $bD_1$, $bD_2$ and $D_1D_2$ planes for sites 1 and 2.} 
\label{gExc}
\end{figure}

The $\mathbf{g}$ tensor  of the \dfc(0) excited state for the $I=0$ isotopes was determined \SW{using optical measurements}, by recording  transmission spectra under a magnetic field of about 200 mT. The low 0-0 inhomogeneous linewidth allowed observing four partially resolved transitions in most orientations for both sites. Fig. \ref{gExc}(a) shows a transmission spectrum for site 1 with $B \parallel D_2$. In this case, the ground state Zeeman splitting is larger than the excited one and the energy difference corresponding to the effective $g$ factors of the ground and excited states are indicated in Figure \ref{gExc}(a). Effective $g$ factors were determined in the three $D_1D_2$, $bD_1$ and $bD_2$ planes by rotating the crystal with respect to the magnetic field (Fig. S2 and S3). \SW{The values for $B\parallel b$ are in agreement with those reported in [\onlinecite{Denoyer:2008kx}].} In a first step, the ground state $g$ tensor \SW{measured by EPR} was compared to the experimental data  to check for sample misalignment. The corresponding angles were then used to fit the excited state Zeeman splittings  to the spin Hamiltonian in the same way as for the ground state. \SW{The calculated effective excited state $g$ factors are shown in Fig. \ref{gExc}(b) and (c).}

Table \ref{magTable} gives the principal values and axes directions of the excited state $g$ tensors for both sites. The largest values are close, 3.4 and 3.3 for sites 1 and 2, and  lower than those of the ground state by a factor of about 2. This can be qualitatively understood by considering the expression of  the $g_{zz}$ component for a pure $M_J$ crystal field level: $g_{zz} = 2g_JM_J$, where $g_J$ is the Land\'e factor. With $M_J=7/2$ and 5/2, we find $g_{zz} = 8$ and 4.3, which reproduces approximately the experimental ratio of 2. The principal axes corresponding to the largest principal $g$ values are oriented close to the $D_1$ and $b$ axes for sites 1 and 2. These orientations are close to those obtained for the ground state, which explains that the angular variations and effective $g$ factors in the $D_1D_2$, $bD_1$ and $bD_2$ planes look similar (Figs. \ref{gFact}, \ref{gExc}(b,c), S2 and S3). 

\begin{table*}%[H] add [H] placement to break table across pages
 \caption{\label{magTable} Magnetic properties of \yb:\YSO. Principal values of the $\mathbf{g}$ and $\mathbf{A}$ ($^{171}$\yb, in MHz)  tensors and Euler angles (in degree) defining the principal axis orientations (zxz convention). }
\begin{ruledtabular}
 \begin{tabular}{lcccccccccccc}
& $g_x$ & $g_y$ & $g_z$ & $\alpha$ & $\beta$ & $\gamma$& $A_x$ & $A_y$ & $A_z$ & $\alpha_A$ & $\beta_A$ & $\gamma_A$ \\
Site 1 \\
Ground state &   -0.31 & -1.60 & 6.53 & 252.8 & 88.7 & 113.8 & 0 & -2140 & -5302 & 247 & 67 & 122  \\   
Excited state  & -0.8 & 1.0 & 3.4 & 77 & 84 & 173\\
 Site 2           \\
Ground state        & -0.13 & -1.50 & 6.06 & 59.1 &11.8 & 347.4   & 2 & 1490 & -4760 & 51 & 11 & 12\\
Excited state        &    -1.0 & 1.4 & -3.3 & 234 & 157 & 190\\
 \end{tabular}
\end{ruledtabular}
 \end{table*}

It was shown in \er:YSO that a long optical coherence lifetime could be obtained with strong fields applied in direction where large effective g factors for the ground and excited states are observed in both sites.\cite{Bottger:2009ik}  At low temperatures, the upper Zeeman levels are strongly depopulated in this configuration, and \er{} spin \SW{flips} are suppressed. Dephasing of the optical transitions between the lowest Zeeman levels are thus strongly decreased too. In \er:YSO, this could be obtained with a field in the $D_1D_2$ plane, for which subsites are equivalent. From Figs. \ref{gFact} and \ref{gExc}, it is clear that a similar strategy in \yb:YSO requires a magnetic field in the $bD_1$ plane, which results in magnetically non equivalent subsites and, for example, a reduced optical density. In the $D_1D_2$ plane or along the $b$ axis, where subsites are magnetically equivalent, one site has always a low effective $g$ factor. Additional experiments are however needed to fully explore this question, as spin \SW{dephasing} mechanisms may have complex dependences with respect to RE concentration, temperature and magnetic field magnitude and orientation.\cite{Bottger:2006jo} 
%\MA{***}Moreover, the large Zeeman splittings that maximize optical coherence lifetime would also  strongly reduce the electron spin ones. Optimal configurations may therefore depend on the application targeted. 

\begin{figure}
\includegraphics[width=0.85\columnwidth]{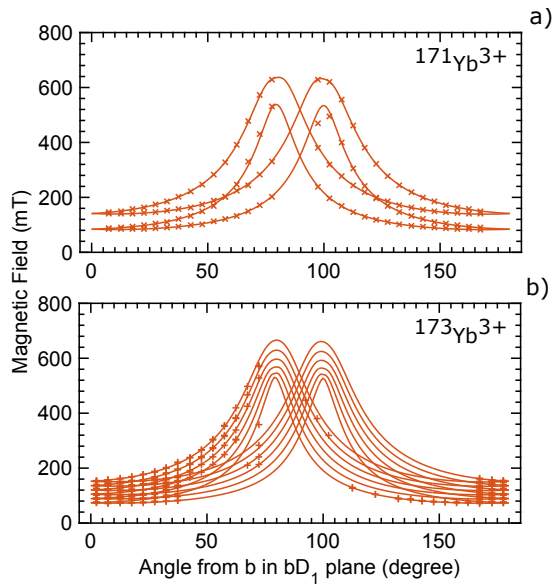}
\caption{Angular variations of the EPR transitions of (a) $^{171}$\yb{} ($I=1/2$) and (b) $^{173}$\yb{} ($I=5/2$) in the $bD_1$ plane in site 2. Crosses: experimental data, lines: fit to a spin Hamiltonian model.} 
\label{aVar}
\end{figure}

We finally determined the ground state hyperfine interaction tensors $\mathbf{A}$ for $^{171}$\yb{} ($I=1/2$) from EPR experiments.  The angular variations of the corresponding EPR lines in the  $D_1D_2$, $bD_1$ and $bD_2$ planes were fitted to the Hamiltonian:
\begin{equation}
H=\mu_B \mathbf{B}\cdot \mathbf{g}\cdot \mathbf{S}+\mathbf{I}\cdot \mathbf{A} \cdot \mathbf{S},
\end{equation}
using the $\mathbf{g}$ tensors and misalignment angles previously determined from the transitions of the $I=0$ isotopes. Fig. \ref{aVar}(a) shows the angular variation  of the two transitions between hyperfine levels in the $bD_1$ plane for the two subsites of site 2. Experimental data are very well reproduced by the fit, as in the case of the other angular variations (Figs. S4).
The principal values and  principal axes of $\mathbf{A}$ are given in Table \ref{magTable} for sites 1 and 2. 

For both sites, the ratio between the tensor elements $A_{ij}/g_{ij}$ is nearly constant for the large $A_{ij}$ and $g_{ij}$ values ($A_{ij}/g_{ij}\approx-0.0264$ for site 1, $A_{ij}/g_{ij}\approx-0.0259$ for site 2), as expected for a pure $J$ multiplet. Indeed, $A_{ij}/g_{ij}=A_J/g_J$ where $A_J$ does not depend on $M_J$. As a result, the $\mathbf{g}$ and $\mathbf{A}$ tensors of both sites have nearly the same orientation as $J$ mixing by the crystal field should be very small given the large separation between the \dfs{} and \dfc{} multiplets (10000 \wnm) compared to the multiplet splittings ($<1000$ \wnm). The larger deviation between the $g$ and $A$ principal axes orientations for site 2 could be due to an enhanced $J$ mixing effect, consistent with the stronger crystal field of the CN = 6 environment (see Section \ref{optSpec}). 
%We can see that the ratio $A_{ij}/g_{ij}$ vary a little bit more for site 2. The reason for this would a not total absence of $J$ mixing by the crystal,  as it was already observed for site 2 in \er:YSO [\onlinecite{GuillotNoel:2006fi}]. 
%

\begin{figure}
\includegraphics[width=0.85\columnwidth]{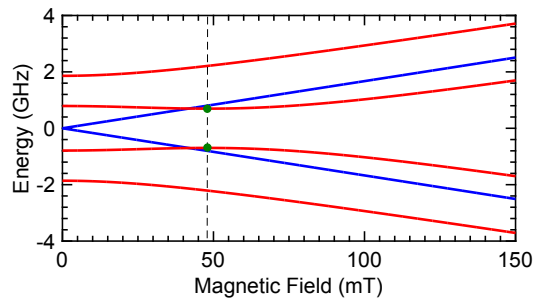}
\caption{Calculated energies $E$ of the ground state hyperfine levels of $^{171}$\yb ($I=1/2$) in site 1 (red lines) as a function of the magnetic field strength. The field is oriented along $D_2$. The vertical dashed line denotes a partial ZEFOZ transition ($dE/dB$=0). The energies of the Zeeman levels of the $I=0$ isotopes are given for comparison (blue lines). } 
\label{aField}
\end{figure}

The large hyperfine interaction results in zero field splittings of 3.7 and 3.1 GHz for sites 1 and 2,\SW{which can} be tuned by several GHz using small magnetic fields oriented along the large effective $g$ factors (Fig. \ref{aField} for site 1 and S7 for site 2). Moreover, ZEFOZ transitions \SW {could} appear when the Zeeman and hyperfine interactions start to be comparable. These transitions, also known as clock transitions, are insensitive to magnetic field fluctuations, which can increase their coherence lifetime.\cite{Wolfowicz:2013ix,Fraval:2004cu} 
A partial ZEFOZ transition is observed around 48 mT in site 1 for a magnetic field oriented along $D_2$ in Fig. \ref{aField}. 
This could be useful for coupling \yb{}  ions to superconducting resonators and obtaining quantum memories for microwave photons with long storage time. 

We also deduced the hyperfine tensor for $^{173}$\yb{} by scaling  $^{171}$\yb{} $\mathbf{A}$ tensor by the ratio (-0.27)  of the nuclear gyromagnetic factors. This accounted for the observed angular variations of $^{173}$\yb{} strongest EPR transitions (Fig. \ref{aVar}(b), S5 and S6), although the quadrupole interaction
induced small energy shifts and extra $\Delta M_I \neq 0$ transitions (Fig. \ref{EPRspec}). The calculated zero field splittings are about 3.5 GHz for both sites and show  complex behaviors at low fields, including many partial ZEFOZ transitions (Fig. S7). However, this should be taken only as an indication since the quadrupole interaction was not included.

\section{Conclusion}

Optical properties of a 50 ppm doped \yb:\YSO{} crystal have been studied in the context of applications in  quantum information processing. In particular, a detailed study of the  transition between the lowest crystal field levels of the \dfs{} and \dfc{} multiplets has been carried out at low temperature, allowing the measurements of  inhomogeneous broadenings $\Gamma_h$, peak absorption coefficients \SW{of} polarized light, oscillator strengths $P$, and excited state lifetimes $T_1$.

 For \yb{} ions in site 2, we found of $P=6.4 \times 10^{-7}$, one of the largest value observed for a rare earth ion in \YSO, while $\Gamma_{inh} = 1.7$ GHz. A relatively strong branching ratio  was also found for the 0-0 transition, leading to a relaxation rate of $T_{1s} = 6.5$ ms and a ratio $T_{1s}/T_1=5.0$. These values also compare favorably with those obtained for other RE doped materials. 
 
 We also determined the ground and excited state Zeeman tensors of the $I=0$ isotopes from angular variations obtained respectively with EPR and optical transmission under a magnetic field of about 200 mT. The largest principal values for both sites are close to $g=6$ for the ground state and $g=3$ for the excited state. The corresponding principal axes are close for the ground and excited states and oriented respectively along $D_1$ and $b$ for sites 1 and 2. 
 
\SW{ \yb{} is the only paramagnetic rare earth ion with a $I=1/2$ isotope, which could be advantageous for optical addressing of spin transitions.}   We therefore determined the ground state $A$ tensor for the $^{171}$\yb{} ($I=1/2$) isotope from EPR measurements. Principal values as large as 5 GHz are observed, leading to calculated total zero field splittings of 3.7 and 3.1 GHz for sites 1 and 2. With magnetic fields oriented along directions of large effective $g$ factors, ground state transitions can be tuned by several GHz with fields of tens of mT. In regions were the Zeeman and hyperfine interactions are comparable in strengths transitions insensitive to magnetic field fluctuations are predicted, which could lead to increased coherence lifetimes. Extrapolated $A$ tensor for  $^{173}$\yb{} ($I=5/2$) isotope is in reasonable agreement with EPR angular variations (Figs. S5 and S6). Calculated zero field splittings are in the same range as for $^{171}$\yb{}, with a much complex behavior under magnetic field. 

In conclusion, these measurements suggest that \yb:\YSO{} could be used for quantum information processing with optical, electron and nuclear spin degrees of freedom.

\begin{acknowledgments}
The authors would like to thank Jean-Fran\c{c}ois Engrand and John G. Bartholomew for their fruitful help. This work received funding from the ANR project DISCRYS (No. 14-CE26-0037-01), Idex no. ANR-10-IDEX-0001-02 PSL and Nano'K project RECTUS.
\end{acknowledgments}

% Create the reference section using BibTeX:
%\bibliography{/Users/philippe/Documents/Biblio/Papers}
%merlin.mbs apsrev4-1.bst 2010-07-25 4.21a (PWD, AO, DPC) hacked
%Control: key (0)
%Control: author (8) initials jnrlst
%Control: editor formatted (1) identically to author
%Control: production of article title (-1) disabled
%Control: page (0) single
%Control: year (1) truncated
%Control: production of eprint (0) enabled
%

\end{document}